\documentclass[reprint,amsmath,amssymb,superscriptaddress,prb]{revtex4-2}

\usepackage{graphicx}
\usepackage{dcolumn}
\usepackage{bm}
\usepackage{color}
\usepackage{hyperref}
\usepackage{subcaption}

\begin{document}

\title{Fragmentation in Coulomb explosion of hydrocarbon molecules}

\author{Samuel S. Taylor}
\affiliation{Department of Physics and Astronomy, Vanderbilt University, Nashville, Tennessee, 37235, USA}

\author{K\'alm\'an Varga}
\email{kalman.varga@vanderbilt.edu}
\affiliation{Department of Physics and Astronomy, Vanderbilt University, Nashville, Tennessee, 37235, USA}

\author{K\'aroly Mogyor\'osi}
\affiliation{ELI ALPS, ELI-HU Non-Profit Ltd, Wolfgang Sander u. 3, H-6728 Szeged, Hungary.}

\author{Viktor Chik\'an}
\affiliation{ELI ALPS, ELI-HU Non-Profit Ltd, Wolfgang Sander u. 3, H-6728 Szeged, Hungary.}
\affiliation{Department of Chemistry, Kansas State University, Manhattan, Kansas 66506-0401, United States}
\affiliation{ASML, 17082 Thornmint Ct, San Diego, CA 92, United States}

\author{Cody Covington}
\affiliation{Department of Chemistry, Austin Peay State University,
Clarksville, USA}

\begin{abstract}
Fragmentation dynamics in the Coulomb explosion of hydrocarbons, specifically methane, 
ethane, propane, and butane, are investigated using time dependent
density functional theory (TDDFT) 
simulations. The goal of this work is to elucidate 
the distribution of fragments generated under laser-driven Coulomb explosion 
conditions. Detailed analysis reveals the types of fragments formed, their respective 
charge states, and the optimal laser intensities required for
achieving various fragmentations. Our results indicate distinct fragmentation 
patterns for each hydrocarbon, correlating with the molecular structure and 
ionization potential. Additionally, we identify the laser parameters that maximize 
fragmentation efficiency, providing valuable insights for experimental setups. 
This research advances our understanding of Coulomb explosion mechanisms 
and offers a foundation for further studies in controlled molecular fragmentation.
\end{abstract}

\maketitle

\section{Introduction}
Coulomb explosion, a process where molecules undergo rapid ionization 
resulting in repulsive interactions between charged fragments, 
has garnered significant attention due to its relevance across 
various scientific and technological domains. 
The advent of ultra-intense femtosecond lasers has made it possible 
to probe this phenomenon with high temporal resolution, 
making it a key subject of both theoretical and experimental 
investigations. Applications of Coulomb explosion span molecular 
imaging \cite{
doi:10.1126/science.244.4903.426,
annurev:/content/journals/10.1146/annurev-physchem-090419-053627,
PhysRevResearch.4.013029,doi:10.1126/science.1246549,Unwin2023,
Mogyorosi2020,doi:10.1126/science.1240362,PhysRevA.91.053424,Boll2022,
PhysRevA.109.052813,10.1063/5.0024833,YATSUHASHI201852,C1CP21345H,
WOS:000791982000001,WOS:000720777300004,PhysRevLett.99.258302,PhysRevA.99.023423,
doi:10.1021/acs.jpclett.8b03726,10.1063/5.0098531,doi:10.1021/acs.jpclett.1c03916,
10.1063/5.0200389,PhysRevA.107.023104,doi:10.1080/23746149.2022.2132182,
WOS:000819572500001,WOS:000947237500006,WOS:000978312900001}, 
structural dynamics
\cite{10.1063/1.5041381,doi:10.1126/science.abc2960,Erattupuzha_2017,
L_2006,PhysRevLett.74.3780,PhysRevA.101.012707,WOS:001171493100004,doi:10.1126/science.adk1833,
PhysRevLett.131.143001,doi:10.1126/science.abc2960,D3CP01740K,
Ekanayake2017,Ekanayake2018}, 
generation of bright keV x-ray photons 
\cite{PhysRevLett.75.3122,McPherson1994}, 
and production of highly energetic electrons 
\cite{PhysRevLett.77.3343}. In addition, fragment formation during 
Coulomb explosion has been extensively studied 
\cite{Hishikawa1998,https://doi.org/10.1002/cphc.201501118,Luzon2019,
D3CP01740K,Bhattacharyya2022,Cornaggia_1992,
C9RA02003A,XU2009255,XU2010119,10.1063/1.3561311,10.1063/1.3561311,Severt2024,
10.1063/1.3561311,doi:10.1021/acs.jpca.3c05442,10.1063/1.5070067}.

Hydrocarbon molecules, particularly methane, ethane, propane, and butane, 
have been used as precursors in numerous Coulomb explosion experiments 
\cite{PhysRevLett.106.163001,PhysRevLett.92.063001,Cornaggia_1992,PhysRevA.51.1431,
SHIMIZU2000609,PhysRevA.82.043433}. These studies have reported key observations such as 
the kinetic energies of ejected protons, emission spectra from the resulting plasma, 
and fragment yields over a range of pulse intensities. However, a comprehensive 
dynamical description of the fragmentation process is often missing\textemdash there 
has been limited focus on identifying the optimal laser pulse parameters and 
intensities that maximize fragmentation efficiency for the first four alkanes in the series.

Previous theoretical efforts to model Coulomb explosions include classical models, which neglect quantum effects~\cite{PhysRevLett.132.123201,PhysRevResearch.4.013029,Unwin2023}, semiclassical methods~\cite{WOS:001290426300001,10.1063/5.0024833}, and \textit{ab initio} approaches~\cite{Corrales2012,PhysRevA.92.053413,OhmuraNagayaShimojoYao+2021+169+187,PhysRevA.95.052701,PhysRevA.91.023422,PhysRevA.89.023429,PhysRevA.86.043407}. Classical models often assume purely repulsive forces between positively charged atomic cores, lacking the attractive forces from electron densities that could form bonds and define molecular fragments. In contrast, \textit{ab initio} methods, such as TDDFT
\cite{PhysRevLett.52.997,ullrich}, capture these electron interactions, 
allowing for the formation of bonded fragments that classical models cannot represent. Given TDDFT's proven accuracy 
in reproducing experimental results~\cite{PhysRevA.92.053413,PhysRevA.95.052701,PhysRevA.91.023422,PhysRevA.89.023429,PhysRevA.86.043407}, we have selected this approach for our study.

In this work, we employ TDDFT simulations to comprehensively examine the fragmentation 
dynamics of hydrocarbons under Coulomb explosion induced by intense laser fields. 
This study is motivated by recent experiments that observed CH radicals and other 
fragments resulting from the Coulomb explosion of ethene \cite{Li2020}
and butane \cite{submitted}. The results are somewhat surprising, 
as earlier experiments \cite{PhysRevA.86.043407,Erattupuzha_2017} primarily demonstrated that strong 
laser pulses predominantly stripped protons from hydrocarbon molecules via a CH bond stretching mechanism.
We will analyze the nature and charge distributions of the resulting fragments, as well as the electron 
dynamics during ionization and the post-ionization molecular motion. Furthermore, 
we will identify the laser intensities that yield optimal fragmentation, providing critical 
insights for designing experiments. 

Our focus on methane, ethane, propane, and butane allows us to uncover distinct fragmentation 
behaviors linked to the molecular structure and ionization potentials of these hydrocarbons. 
By pinpointing the laser parameters that maximize fragmentation efficiency, this study 
enhances the understanding of Coulomb explosion mechanisms and lays the groundwork for 
future investigations into controlled molecular fragmentation.

\section{Computational Method}
The simulations were performed using TDDFT  for modeling the electron dynamics on a 
real-space grid with real-time propagation \cite{Varga_Driscoll_2011a}, 
with the Kohn-Sham (KS) Hamiltonian of the following form
\begin{equation}
\begin{split}
\hat{H}_{\text{KS}}(t) = -\frac{\hbar^2}{2m} \nabla^2 + V_{\text{ion}}(\mathbf{r},t) 
+ V_{\text{H}}[\rho](\mathbf{r},t) \\
+ V_{\text{XC}}[\rho](\mathbf{r},t) + V_{\text{laser}}(\mathbf{r},t).
\end{split}
\label{eq:hamiltonian}
\end{equation}
Here, $\rho$ is the electron density which is defined as the density sum over all occupied orbitals:
\begin{equation}
\rho(\mathbf{r},t) = \sum_{k=1}^{N_{\text{orbitals}}} 2|\psi_k(\mathbf{r},t)|^2,
\end{equation}
where the coefficient 2 accounts for there being two electrons in each orbital (via spin degeneracy) and $k$ is a quantum number labeling each orbital. 

$V_{ion}$ in Eq.~\eqref{eq:hamiltonian} is the external potential due to the ions, represented by employing norm-conserving pseudopotentials centered at each ion as given by Troullier and Martins~\cite{PhysRevB.43.1993}. $V_{H}$ is the Hartree potential, defined as
\begin{equation}
V_H(\mathbf{r}, t) = \int \frac{\rho(\mathbf{r}', t)}{|\mathbf{r} - \mathbf{r}'|} \, d\mathbf{r}',
\end{equation}
and accounts for the electrostatic Coulomb interactions between
electrons. The term $V_{XC}$ is the exchange-correlation potential,
which is approximated by the adiabatic local-density approximation
(ALDA), obtained from a parameterization to a homogeneous electron gas by Perdew and Zunger~\cite{PhysRevB.23.5048}. The last term in Eq.~\eqref{eq:hamiltonian}, $V_{laser}$ is the  time–dependent potential due to the electric field of the laser, and is described using the dipole approximation, $V_{\text{laser}} = \mathbf{r} \cdot \mathbf{E}_{\text{laser}}(t)$. The electric field $\mathbf{E}_{\text{laser}}(t)$ is given by
\begin{equation}
\mathbf{E}_{\text{laser}}(t) = E_{\text{max}} \exp\left[-\frac{(t - t_0)^2}{2a^2}\right] \sin(\omega t) \mathbf{\hat{k}},
\label{eq:laser}
\end{equation}
where the parameters $E_{\text{max}}$, $t_{0}$, and $a$ define the maximum amplitude, initial position of the center, and the width of the Gaussian envelope, respectively. $\omega$ describes the frequency of the laser and $\mathbf{\hat{k}}$ is the unit vector in the polarization direction of the electric field.

At the beginning of the TDDFT calculations, the ground state of the system is prepared by performing a Density-Functional Theory (DFT) calculation. With these initial conditions in place, we then proceed to propagate the Kohn–Sham orbitals, $\psi_{k}(\mathbf{r},t)$ over time by using the time-dependent KS equation, given as 
\begin{equation}
i \frac{\partial \psi_k(\mathbf{r}, t)}{\partial t} = \hat{H} \psi_k(\mathbf{r}, t).
\label{eq:tdks}
\end{equation}
Eq.~\eqref{eq:tdks} was solved using the following time propagator
\begin{equation}
\psi_k(\mathbf{r}, t + \delta t) = \exp\left(-\frac{i \hat{H}_{\text{KS}}(t) \delta t}{\hbar}\right) \psi_k(\mathbf{r}, t).
\end{equation}
This operator is approximated using a fourth-degree Taylor expansion, given as
\begin{equation}
\psi_k(\mathbf{r}, t + \delta t) \approx \sum_{n=0}^{4} \frac{1}{n!} \left(\frac{-i \delta t}{\hbar} \hat{H}_{\text{KS}}(t)\right)^n \psi_k(\mathbf{r}, t).
\end{equation}
The operator is applied for $N$ time steps until the final time, $t_{final} = N \cdot \delta t$, is obtained. A time step of $\delta t = 1$ as was used in the simulations.

In real-space TDDFT, the Kohn-Sham orbitals are represented at discrete points in real space. These points are organized on a uniform rectangular grid. The accuracy of the simulations is determined by the grid spacing, which is the key parameter that can be adjusted. In our simulations, we used a grid spacing of 0.3 Å and placed 100 points along each of the $x$, $y$, and $z$ axes.

To enforce boundary conditions, we set the Kohn-Sham orbitals to zero at the edges of the simulation cell. However, when a strong laser field is applied, ionization can occur, potentially causing unphysical reflections of the wavefunction at the cell boundaries. To address this issue, we implemented a complex absorbing potential (CAP) to dampen the wavefunction as it reaches the boundaries. The specific form of the CAP used in our simulations, as described by Manopolous~\cite{10.1063/1.1517042}, is given by:
\begin{equation}
- i w(x) = -i \frac{\hbar^2}{2m} \left(\frac{2\pi}{\Delta x}\right)^2 f(y),
\end{equation}
where $x_{1}$ is the start and $x_{2}$ is the end of the absorbing region, $\Delta x = x_{2} - x_{1}$, $c = 2.62$ is a numerical constant, $m$ is the electron’s mass, and
\begin{equation}
f(y) = \frac{4}{c^2} \left( \frac{1}{(1 + y)^2} + \frac{1}{(1 - y)^2} - 2 \right), \quad y = \frac{x - x_1}{\Delta x}.
\end{equation}

As the molecule is ionized by the laser field, the electron density is directed towards the CAP. Additionally, the ejected fragments carry their electron density move towards the CAP. When any electron density into contact with the CAP, it is absorbed. Consequently, the total electron number
\begin{equation}
N(t) = \int_V \rho(\mathbf{r}, t) \, d^3x,
\end{equation}
where $V$ is the volume of the simulation box, will diverge from the initial electron number, $N(0)$. We interpret $N(0) - N(t)$ as the total number of electrons ejected from the simulation box.

Motion of the ions in the simulations were treated classically. Using the Ehrenfest theorem
, the quantum forces on the ions due to the electrons are given by the derivatives 
of the expectation value of the total electronic energy with respect to the ionic positions. 
These forces are then fed into Newton’s Second Law, giving
\begin{equation}
\begin{split}
M_i \frac{d^2 \mathbf{R}_i}{dt^2} = Z_i \mathbf{E}_{\text{laser}}(t) 
+ \sum_{j \neq i}^{N_{\text{ions}}} \frac{Z_i Z_j (\mathbf{R}_i - \mathbf{R}_j)}{|\mathbf{R}_i - \mathbf{R}_j|^3}\\ 
- \nabla_{\mathbf{R}_i} \int V_{\text{ion}}(\mathbf{r}, \mathbf{R}_i) \rho(\mathbf{r}, t) \, d\mathbf{r},
\end{split}
\end{equation}
where $M_{i}$, $Z_{i}$, and $\mathbf{R}_{i}$ are the mass, pseudocharge (valence), and position of the $i$-th ion, respectively, and $N_{\text{ions}}$ is the total number of ions. This differential equation was time propagated using the Verlet algorithm at every time step $\delta t$. 
This approach has been successfully used to describe the Coulomb explosion of molecules~\cite{PhysRevA.89.023429,
PhysRevA.91.023422,PhysRevA.92.053413,PhysRevA.95.052701,PhysRevA.86.043407}.

In each simulation, the ion velocities are initialized using a Boltzmann distribution corresponding to 300 K. This random initialization facilitates the exploration of various fragmentation pathways during the Coulomb explosion. If the ion velocities were held constant, the simulations would yield identical fragmentation outcomes, thus constraining the statistical variability of the results. To ensure a comprehensive distribution of data, more than 
50 simulations were conducted for each molecule, 
allowing for the formation of C$_n$H$_m$ fragments in several instances across all tested molecules. Due to limited computational resources, no further simulations were performed.

Each molecule in the simulations was positioned at the center of the simulation box, with its longest molecular axis aligned along the \(x\)-axis and 
its shortest axis oriented along the \(z\)-axis. In every simulation, the electric field was polarized along the \(x\)-axis to maximize ionization~\cite{PhysRevA.92.053413}. These conditions model experiment, where precise molecular alignment is achievable 
\cite{annurev:/content/journals/10.1146/annurev-physchem-090419-053627}.

In the following section, we present the results from a substantial
number of simulations conducted on the first four alkanes: methane
(CH\textsubscript{4}), ethane (C\textsubscript{2}H\textsubscript{6}), propane (C\textsubscript{3}H\textsubscript{8}), and butane (C\textsubscript{4}H\textsubscript{10}), using a laser intensity determined to optimize fragmentation. The laser parameters, including wavelength and duration, were modeled after the pulse described in \cite{submitted}, which follows the functional form outlined in Eq.~\eqref{eq:laser}. The only varying parameter in our simulations was the laser intensity, which was adjusted to explore the ionization, fragmentation, and dissociation ranges for the selected alkanes.

\subsection{Methane (CH\textsubscript{4})}

\begin{figure*}
\centering
\includegraphics[width=\textwidth]{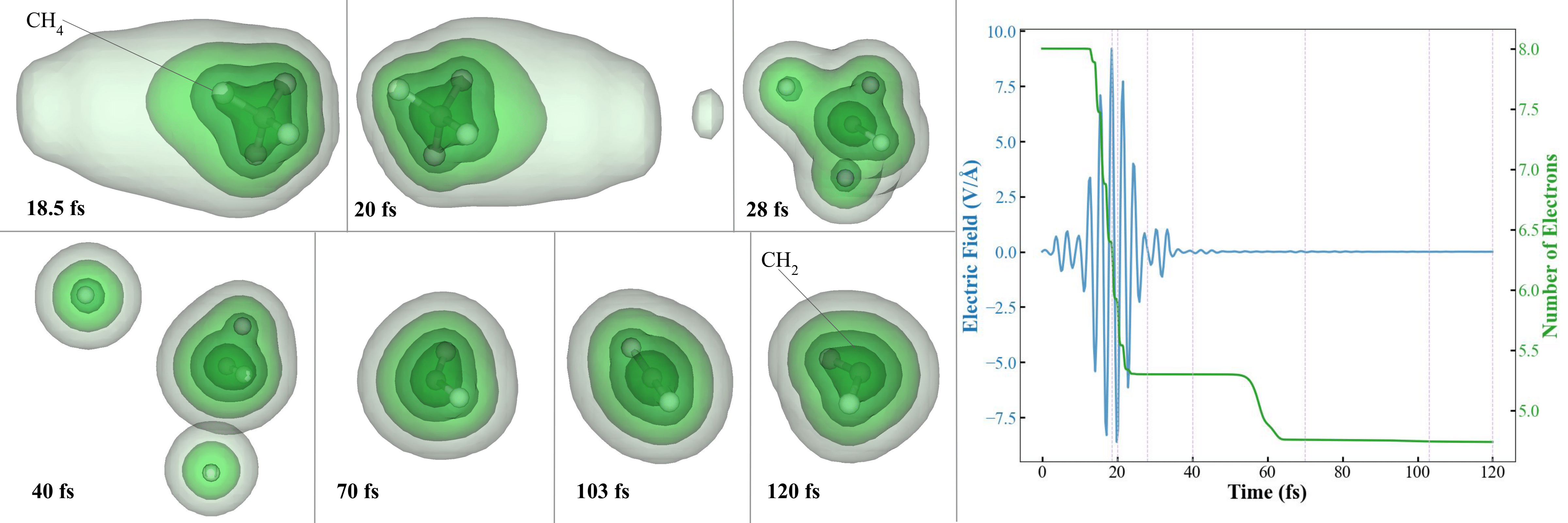}
\caption{CH\textsubscript{4} Coulomb explosion snapshots resulting in the formation of 
one CH\textsubscript{2} molecule and two discharged H atoms  have  $1.26^+$, $0.62^+$, and $0.84^+$ charges, 
respectively (the 0.5, 0.1, 0.01, and 0.001 density isosurfaces are displayed). 
The pulse electric field and the number of electrons in the simulation are shown, 
with vertical dashed purple lines indicating the specific times in the simulation when each snapshot was captured.}
\label{CH4-snapshot}
\end{figure*}

\begin{figure*}
\centering
\includegraphics[width=.9\textwidth]{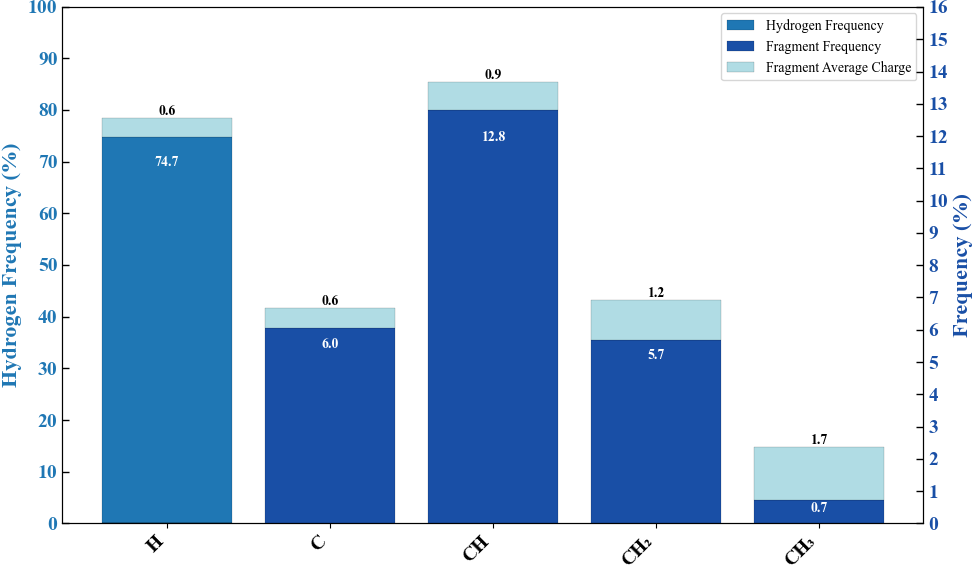}\\
\caption{Average production frequency of each fragment generated in 71
CH$_4$ Coulomb explosion simulations. 
The darker columns represent the formation frequency of hydrogen fragments and 
carbon-containing fragments. The lighter appended columns indicate the corresponding average charges associated with each 
produced fragment}
\label{CH4-histogram}
\end{figure*}

\begin{figure*}
\centering
\includegraphics[width=\textwidth]{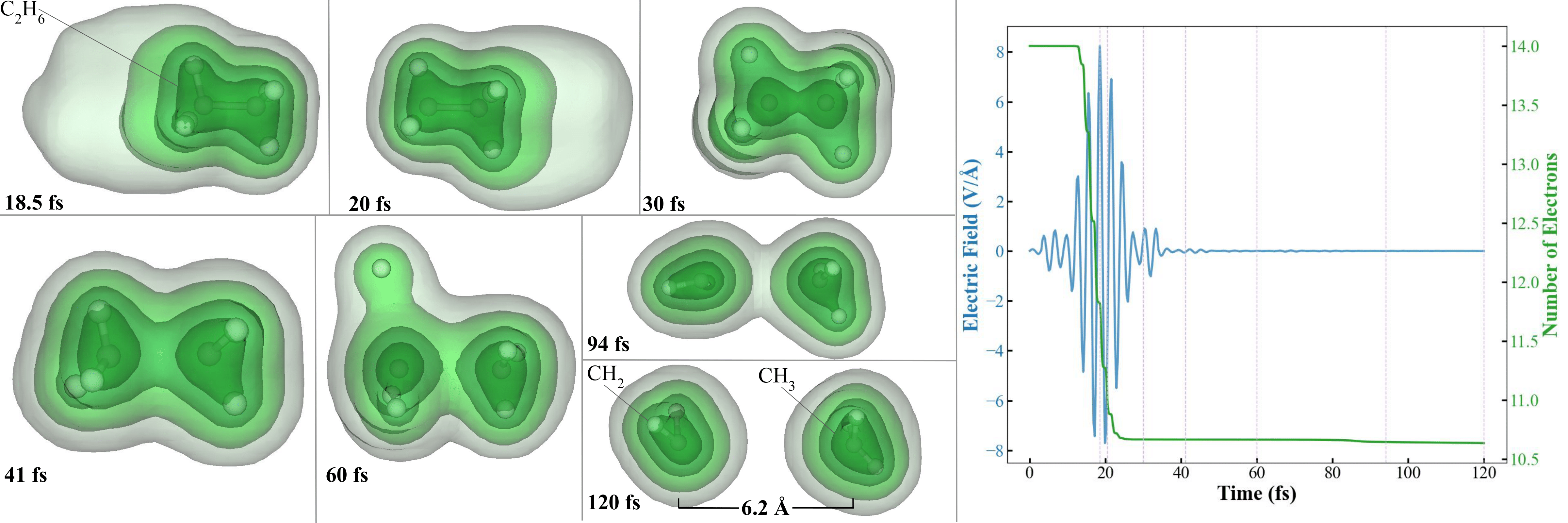}
\caption{Snapshots of the Coulomb explosion of C\textsubscript{2}H\textsubscript{6}, 
illustrating the formation of CH\textsubscript{2} and CH\textsubscript{3} fragments, 
which carry charges of \(1.18^{+}\) and \(1.18^{+}\), respectively (the 0.5, 0.1, 0.01, and 0.001 
density isosurfaces are shown). The bar between the two fragments at 120 fs denotes the distance between their center of masses.}
\label{C2H6-snapshot}
\end{figure*}

\begin{figure*}
\centering
\includegraphics[width=.9\textwidth]{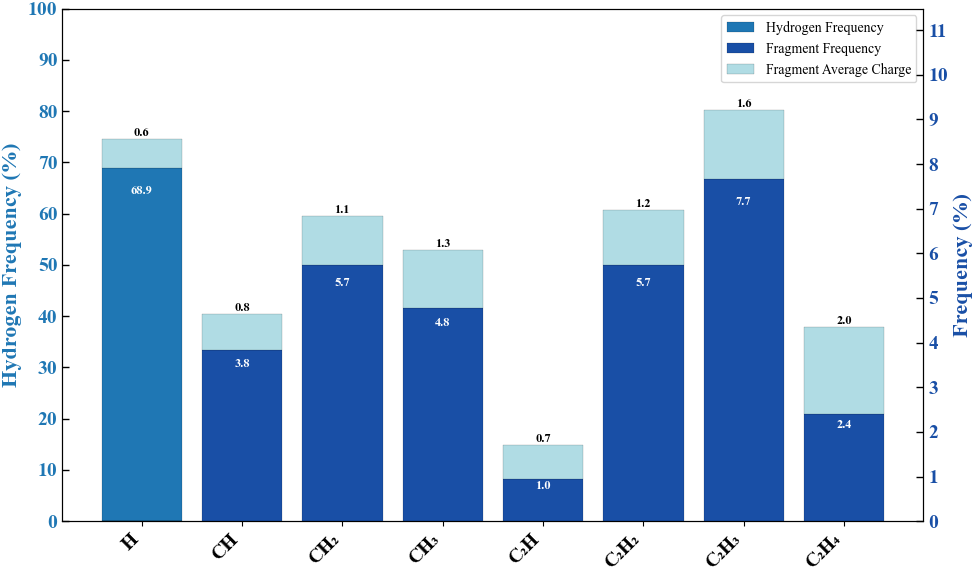}\\
\caption{Histogram illustrating the average production frequency and charge of each 
fragment generated in 50 C\textsubscript{2}H\textsubscript{6} Coulomb explosion simulations.}
\label{C2H6-histogram}
\end{figure*}

\begin{table}[b] 
\begin{ruledtabular}
\begin{tabular}{lccc}
\textrm{} &
\textrm{Ionization} &
\textrm{Fragmentation} &
\textrm{Dissociation} \\
\textrm{Alkane} &
\textrm{threshold} &
\textrm{threshold} &
\textrm{threshold} \\
\textrm{} &
\textrm{$(\text{W}/\text{cm}^2)$} &
\textrm{$(\text{W}/\text{cm}^2)$} &
\textrm{$(\text{W}/\text{cm}^2)$} \\
\colrule
CH\textsubscript{4} & $7.0 \times 10^{14}$ & $1.1 \times 10^{15}$ & $1.5 \times 10^{15}$ \\
C\textsubscript{2}H\textsubscript{6} & $6.1 \times 10^{14}$ & $9.0 \times 10^{14}$ & $1.5 \times 10^{15}$ \\
C\textsubscript{3}H\textsubscript{8} & $3.8 \times 10^{14}$ & $7.0 \times 10^{14}$ & $1.5 \times 10^{15}$ \\
C\textsubscript{4}H\textsubscript{10} & $4.5 \times 10^{14}$ & $7.0 \times 10^{14}$ & $1.3 \times 10^{15}$ \\
\end{tabular}
\end{ruledtabular}
\caption{\label{table_laser_parameters}%
Approximate pulse intensities required for ionization, fragmentation, and dissociation of the first four alkanes are presented, with a pulse wavelength of 890 nm and a duration of 7.8 fs (FWHM). Pulse intensities below the ionization threshold result in minimal or no dissociation of the molecular structure. Intensities near the fragmentation threshold cause partial molecular breakup, leading to the formation of smaller ionized fragments. Intensities above the dissociation threshold result in nearly complete molecular disintegration, with all atoms being ejected from the molecule.}

\end{table}
The electron densities and molecular dynamics of a CH\textsubscript{4} molecule during a Coulomb explosion simulation 
are illustrated in Fig.~\ref{CH4-snapshot}. These snapshots are from one of the 71 simulations conducted 
using CH\textsubscript{4} as the precursor molecule. Initially, the snapshots show how the laser 
field rips the electron density away from the molecule, triggering rapid ionization. As shown in the 
figure, significant ionization begins when the electric field magnitude reaches approximately 
$5.5~\text{V}/\text{\AA}$ at 14 fs. The ionization process concludes when the laser field 
falls below $6~\text{V}/\text{\AA}$ at around 24 fs, leaving the molecule with 5.3 valence electrons. 
This near-triple ionization creates strong Coulomb repulsion between the resulting fragments, causing them 
to separate rapidly. In our calculations, the electron density is integrated over a finite volume, resulting in non-integer electron counts. This phenomenon can be interpreted in several ways. One interpretation suggests that approximately 5.3 electrons remain localized within the molecular region during the simulation, with the fractional charge either recombining with the ionized electron cloud or dissociating over a longer simulation period. Alternatively, the non-integer charge could be understood as an average, where some molecular fragments retain 5 valence electrons, others 6, and 5.3 represents the mean electron count per fragment. For example, if a hydrogen atom was ejected with a charge of $0.6^{+}$, that would indicate that 60\% of the time it would be detected as a $1^{+}$ ion, and 40\% of the time it would be a neutral hydrogen atom.

The ejected hydrogen fragments are observed moving toward the CAP, 
where their electrons are absorbed, leading to a further
(artificial) reduction in the number of electrons. After the ionization, the remaining 
fragment (CH\textsubscript{2}) undergoes structural rearrangement. Fig.~\ref{CH4-snapshot}
depicts the CH\textsubscript{2} fragment transitioning from a right-bent geometry at 70 fs to a linear 
conformation at 103 fs, and finally to a left-bent structure at 120 fs.
The laser pulse used in the Coulomb explosion simulations of 
CH\textsubscript{4} (also shown Fig.~\ref{CH4-snapshot}) had an intensity of 
$1.1 \times 10^{15}~\text{W}/\text{cm}^2$, a peak electric field of 
$9.2~\text{V}/\text{\AA}$, a wavelength of $890~\text{nm}$, and a pulse duration of 7.8 fs (FWHM). 
This intensity was found to be optimal for inducing fragmentation of CH\textsubscript{4}, as lower intensities 
(approximately $7 \times 10^{14}~\text{W}/\text{cm}^2$ and below) led to ionization without significant fragmentation, 
while higher intensities (approximately $1.5 \times 10^{15}~\text{W}/\text{cm}^2$ and above) caused complete molecular dissociation 
(see Table~\ref{table_laser_parameters}). Intermediate laser intensities, 
ranging from the ionization to the dissociation thresholds, were also explored. 
At intensities between the ionization and fragmentation threshold, 
fragmentation was less effective, often leading to partial dissociation and the formation of larger fragments, such as CH\textsubscript{3}. 
At intensities between the fragmentation and dissociation threshold, smaller fragments, such as C and CH, were observed with little diversity. 
It is important to note that in experimental conditions, only part of the target is exposed to the highest intensity of the laser pulse, 
while molecules outside the focal region interact with lower intensity
fields. As a result, various fragmentation processes 
can occur simultaneously in experiments.

Fig.~\ref{CH4-histogram} illustrates the distribution of fragments and their corresponding charges resulting 
from the Coulomb explosion of CH\textsubscript{4}. The data demonstrate that the reaction yields a range of 
hydrocarbons smaller than CH\textsubscript{4}. Among these fragments, CH is the most frequently 
observed carbon-containing species, appearing in 12.8\% of the simulations with an average charge 
of 0.9\textsuperscript{+}. Hydrogen fragments are the most commonly detected overall, with an average 
charge of 0.6\textsuperscript{+} (equivalent to 0.4 electrons) as they approach the CAP. Other fragments, such as CH, CH\textsubscript{2}, and CH\textsubscript{3}, are also 
present but occur less frequently. Importantly, no CH\textsubscript{4} fragments were observed, indicating 
that the laser pulse intensity was sufficiently strong to dissociate at least one atom from the molecule 
in every simulation. On average, the laser ejected 2.72 electrons per simulation. These simulations help determine 
the optimal pulse parameters and predict the dissociation pathways for CH\textsubscript{4} in 
Coulomb explosion fragmentation.

\subsection{Ethane (C\textsubscript{2}H\textsubscript{6})}

Fig.~\ref{C2H6-snapshot} presents snapshots from one of the 50 Coulomb explosion 
simulations with C\textsubscript{2}H\textsubscript{6} as the precursor molecule. These snapshots 
illustrate the interaction between the electric field of the laser pulse and the electron 
density of the molecule. The field stretches and compresses the electron density. As shown in the 
pulse diagram in Fig.~\ref{C2H6-snapshot}, when the electric field exceeds $5~\text{V}/\text{\AA}$, between 14 fs 
and 22 fs, rapid ionization occurs, reducing the valence electron count to approximately 10.7. 
Following the dissipation of the laser pulse at 40 fs, two CH$_3$ fragments are
produced. By 60 fs, due to the asymmetric charge distribution, one of the fragments ejects a hydrogen atom.
The ionization induces strong Coulomb repulsion between the 
resulting fragments, causing the CH\textsubscript{2} and CH\textsubscript{3} groups to visibly repel 
each other throughout the simulation.
Additionally, the snapshots reveal rotational motion of the fragments, particularly within 
the CH\textsubscript{2} fragment. Between 60 and 94 fs, the CH\textsubscript{2} molecule rotates 
approximately 90 degrees about the $z$-axis, followed by an additional 90-degree rotation from 94 to 120 fs.

\begin{figure}
    \centering
    \begin{subfigure}[b]{0.45\textwidth}
        \centering
        \includegraphics[width=\textwidth]{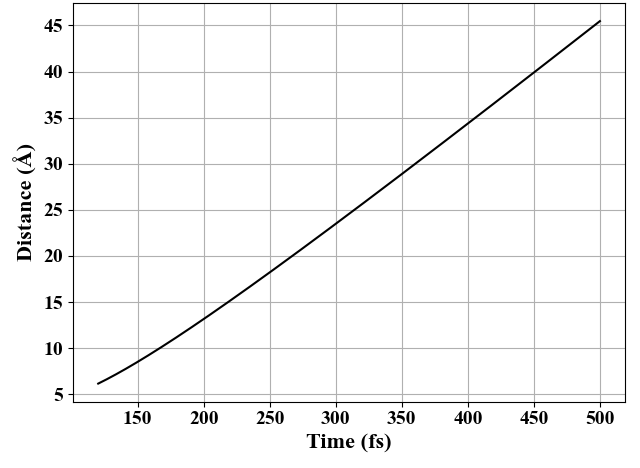}
    \end{subfigure}
    \hfill
    \begin{subfigure}[b]{0.45\textwidth}
        \centering
        \includegraphics[width=\textwidth]{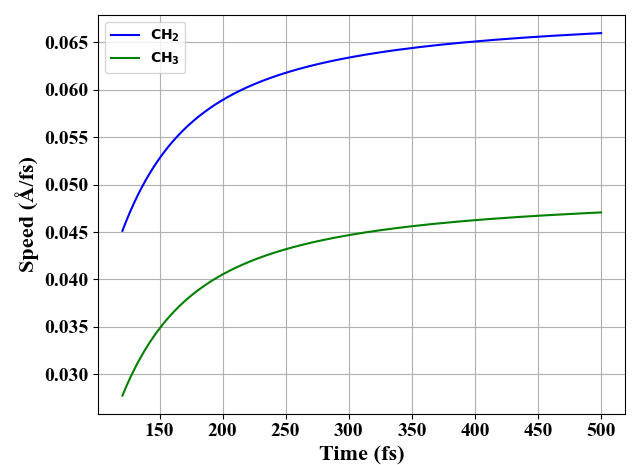}
    \end{subfigure}
    \caption{Distance (top) and speed (bottom) between the CH\textsubscript{2} and CH\textsubscript{3} 
    shown in Fig.~\ref{C2H6-snapshot} from 120 fs to 500 fs. Note that
    1\AA/fs is 100000 m/s.}
    \label{C2H6-trajectory}
\end{figure}

The CH\textsubscript{2}$^+$ and CH\textsubscript{3}$^+$ fragments are well
separated at the end of the TDDFT simulation in Fig.
\ref{C2H6-snapshot}. Fig.~\ref{C2H6-trajectory} shows how the fragments
move after 120 fs assuming that the two charged fragments interact via
the Coulomb potential and their motion can be described by solving the
Newton equations of motions using the Verlet algorithm.
By the conclusion of the TDDFT simulation, the centers of mass of the molecules 
were separated by 6.2 \AA. 
Fig.~\ref{C2H6-trajectory} shows that the acceleration
gradually decreases and the distance between the ions grows linearly. 
At 500 fs, the distance between the molecules increased to 45 \AA, with 
the CH\textsubscript{2}$^+$ and CH\textsubscript{3}$^+$ groups each exhibiting velocities 
of $6.6 \times 10^{-2}$~\AA/fs (6600 m/s) and $4.7 \times
10^{-2}$~\AA/fs (4700 m/s), respectively. The large difference in
velocity is due to the difference between the initial velocities of the
fragments. In experiment, the velocity of the 
CH\textsubscript{2}$^+$ ion in Coulomb explosion of methanol has been
measured to be 4800 m/s \cite{C9RA02003A}, which is within the same magnitude as in the present calculations. Considering that the laser intensity used in the simulations is six times higher than in the experiment, this agreement is remarkably good.

As depicted in Fig.~\ref{C2H6-snapshot}, the laser pulse employed in the Coulomb explosion simulations 
of C\textsubscript{2}H\textsubscript{6} had an intensity of $9 \times 10^{14}~\text{W}/\text{cm}^2$,
a maximum electric field of $8.23~\text{V}/\text{\AA}$, a wavelength of $890~\text{nm}$, and a pulse 
duration of 7.8 fs (FWHM). This intensity was found to be optimal for inducing the richest diversity in fragmentation pathways 
of C\textsubscript{2}H\textsubscript{6}. To evaluate the effects of varying laser intensities, additional simulations were conducted across a broader range, extending beyond the ionization, fragmentation, and dissociation thresholds listed in Table~\ref{table_laser_parameters}. It was determined that $9 \times 10^{14}~\text{W}/\text{cm}^2$ 
achieved the most effective fragmentation because lower intensities led to weak or incomplete dissociation 
(producing CH\textsubscript{3} or not dissociating at all) and higher intensities led to over-dissociation 
(producing only C\textsubscript{2} or complete fragmentation). Consequently, a pulse intensity of $9 \times 10^{14}~\text{W}/\text{cm}^2$ was 
selected for the 50 simulations, yielding the most diverse range of fragmentation pathways for 
C\textsubscript{2}H\textsubscript{6}.

Fig.~\ref{C2H6-histogram} shows the distribution of fragments and their corresponding charges resulting 
from the Coulomb explosion of C\textsubscript{2}H\textsubscript{6}. The data reveal that the fragmentation 
produces various hydrocarbons smaller than C\textsubscript{2}H\textsubscript{6}. Among these fragments, 
C\textsubscript{2}H\textsubscript{3} is the most frequently detected carbon-containing species, 
appearing in 7.7\% of the simulations with an average charge of $1.6^{+}$. Hydrogen fragments are the most 
prevalent overall, carrying an average charge of $0.6^{+}$ as they are ejected from the molecule. 
Other fragments, including CH, CH\textsubscript{2}, and C\textsubscript{2}H\textsubscript{3}, are also observed but with lower 
frequencies. Notably, no C\textsubscript{2}H\textsubscript{6} or C\textsubscript{2}H\textsubscript{5} 
fragments were detected, indicating that the laser pulse intensity was sufficient to dissociate at least two 
hydrogen atoms from the molecule in every simulation. On average, the laser ejected 3.42 electrons per simulation.

\subsection{Propane (C\textsubscript{3}H\textsubscript{8})}

\begin{figure*}
\centering
\includegraphics[width=\textwidth]{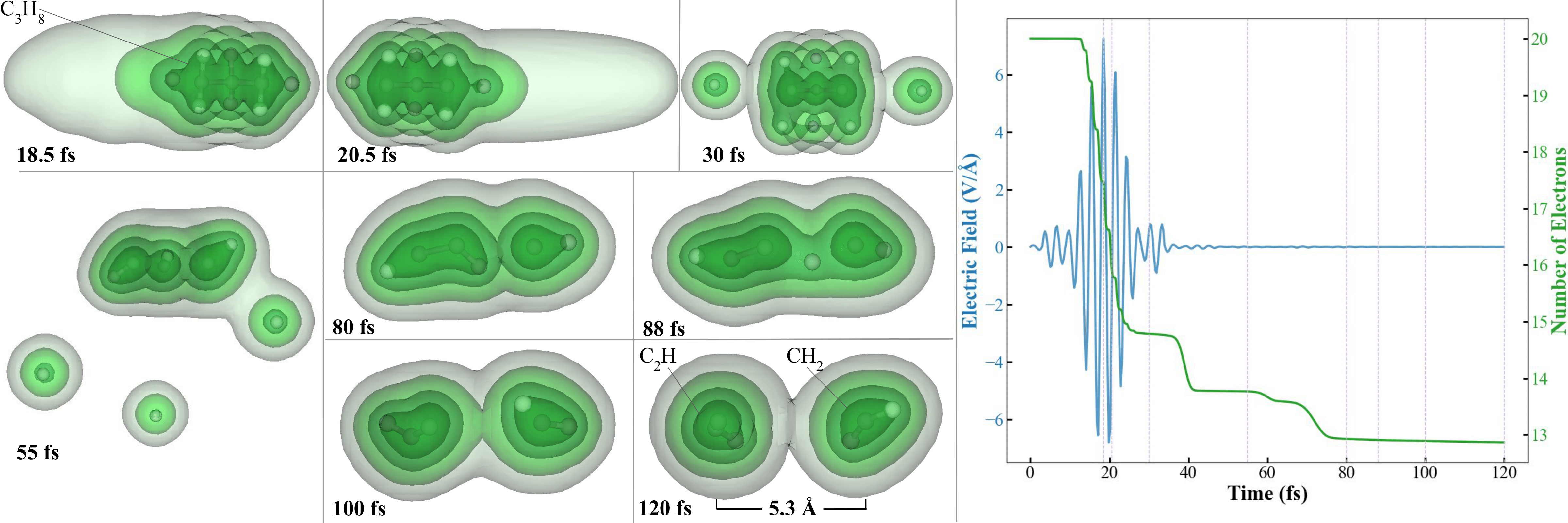}
\caption{C\textsubscript{3}H\textsubscript{8} Coulomb explosion snapshots 
resulting in the formation of C\textsubscript{2}H and CH\textsubscript{2} fragments 
with $2.22^+$ and $0.91^+$ charges, respectively (the 0.5, 0.1, 0.01, and 0.001 density 
isosurfaces are shown). The bar
between the two fragments at 120 fs denotes the distance between their center of masses.}
\label{C3H8-snapshot}
\end{figure*}

\begin{figure*}
\centering
\includegraphics[width=.9\textwidth]{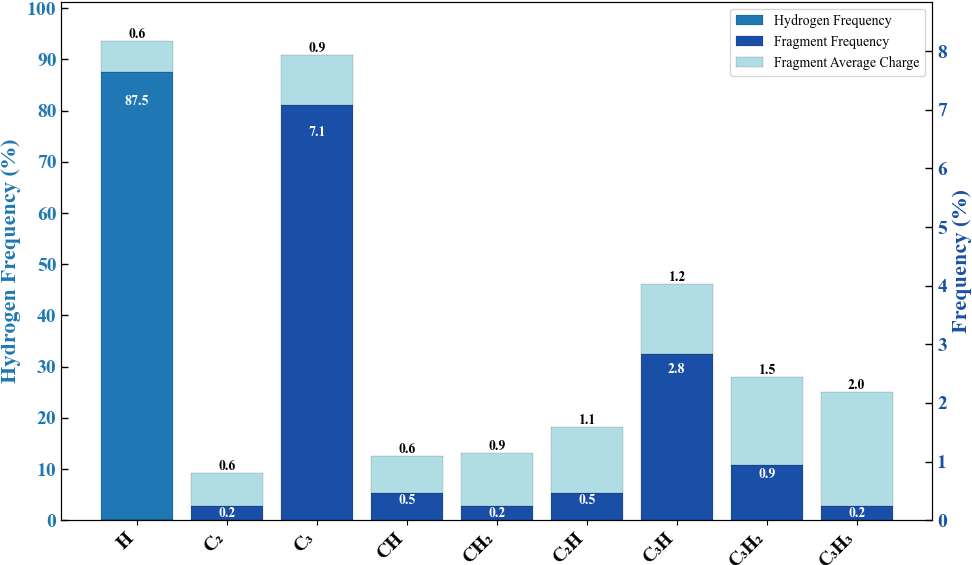}\\
\caption{Histogram illustrating the average production frequency and charge of each fragment generated 
in 50 C\textsubscript{3}H\textsubscript{8} Coulomb explosion simulations.}
\label{C3H8-histogram}
\end{figure*}

Fig.~\ref{C3H8-snapshot} shows snapshots from one of the 50 Coulomb
explosion simulations using C\textsubscript{3}H\textsubscript{8} as
the precursor molecule. These images depict how the laser pulse’s
electric field interacts with the molecule’s electron density,
stretching it and expelling it from the molecule, which leads to fragmentation. 
According to the pulse diagram in Fig.~\ref{C3H8-snapshot}, rapid ionization occurs when the electric field intensity exceeds $4~\text{V}/\text{\AA}$ from approximately 14 fs to 24 fs, resulting in the retention of 14.8 valence electrons in the molecule. This ionization leads to the ejection of two hydrogen atoms at 30 fs, followed by the ejection of three additional hydrogen atoms at 55 fs. At 88 fs in the simulation, a hydrogen atom is observed transferring from the C\textsubscript{2}H fragment to the CH fragment, indicating that molecular rearrangement and bonding can occur even after the laser-induced ionization.
As illustrated in the figure, the resulting Coulomb explosion and subsequent molecular dynamics lead to the formation of C\textsubscript{2}H and CH\textsubscript{2} fragments. The laser pulse employed in the Coulomb explosion simulations of C\textsubscript{3}H\textsubscript{8} featured an intensity of \(7 \times 10^{14}~\text{W}/\text{cm}^2\), a maximum electric field of \(7.26~\text{V}/\text{\AA}\), a wavelength of \(890~\text{nm}\), and a pulse duration of 7.8 fs (FWHM) (refer to Table~\ref{table_laser_parameters} for details regarding the fragmentation pulse intensity threshold).


Fig.~\ref{C3H8-histogram} presents the distribution of fragments and their corresponding charges 
resulting from the Coulomb explosion of C\textsubscript{3}H\textsubscript{8}. The data reveals the diverse set of fragmentation pathways achievable with C\textsubscript{3}H\textsubscript{8} as the precursor molecule, among which C\textsubscript{3}H is the most commonly observed hydrocarbon, appearing in 2.8\% of the simulations with an average charge of $1.2^{+}$.
Hydrogen fragments are the most frequently detected overall, 
exhibiting an average charge of $0.6^{+}$ as they near the CAP. Other fragments, such as 
CH, C\textsubscript{3}, and C\textsubscript{3}H, are also present but with a lower formation rate. Importantly, no fragments 
of C\textsubscript{3}H\textsubscript{4} or larger were found, suggesting that the laser pulse intensity was 
adequate to remove at least five hydrogen atoms from the molecule in each simulation. On average, the laser 
expelled 5.15 electrons per simulation. 

\subsection{Butane (C\textsubscript{4}H\textsubscript{10})}

\begin{figure*}
\centering
\includegraphics[width=\textwidth]{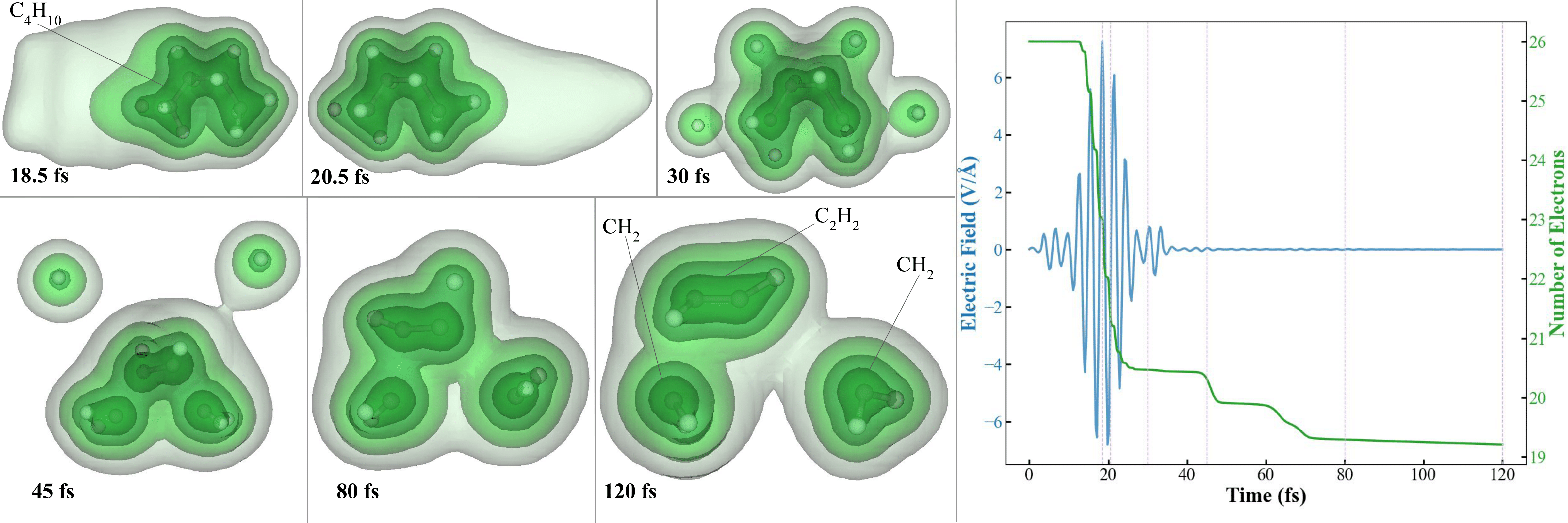}
\caption{C\textsubscript{4}H\textsubscript{10} Coulomb explosion snapshots resulting in the 
formation of C\textsubscript{2}H\textsubscript{2} and two CH\textsubscript{2} fragments 
measured to have $1.12^+$, $0.77^+$ and $0.90^+$ charges, 
respectively (the 0.5, 0.1, 0.01, and 0.001 density isosurfaces are shown).}
\label{C4H10-snapshot}
\end{figure*}

\begin{figure*}
\centering
\includegraphics[width=.9\textwidth]{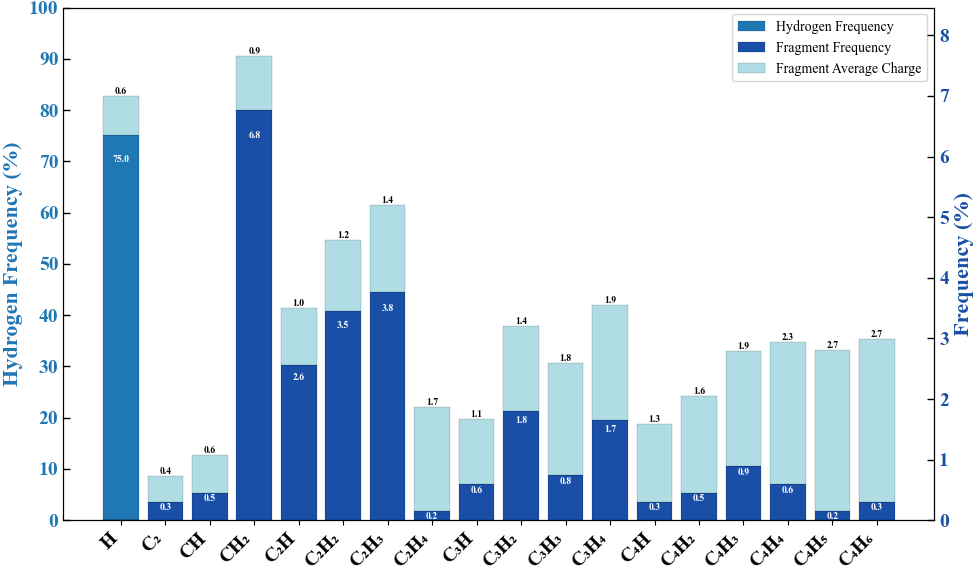}\\
\caption{Histogram illustrating the average production frequency and charge of each fragment generated 
in 88 C\textsubscript{4}H\textsubscript{10} Coulomb explosion simulations.}
\label{C4H10-histogram}
\end{figure*}

Fig.~\ref{C4H10-snapshot} shows snapshots from one of the 88 Coulomb explosion simulations conducted with C\textsubscript{4}H\textsubscript{10} in the gauche conformation the precursor molecule These snapshots capture the dynamic 
interaction between the laser pulse's electric field and the electron density of the molecule, which causes 
the electron density to stretch and be expelled toward the CAP. As illustrated in 
the pulse diagram in Fig.~\ref{C4H10-snapshot}, when the electric field intensity surpasses $4~\text{V}/\text{\AA}$ 
(between 14 fs and 24 fs), a rapid ionization process occurs, leaving 20.5 valence electrons in the molecule. 
Similarly to the previous case, by the end of the laser pulse at approximately 30 fs, two hydrogen atoms are ejected. 
Additionally, the CH bond involving the two hydrogen atoms at the top is significantly stretched, resulting in further ejection around 45 fs.
In this instance, the system ultimately forms a charged C\textsubscript{2}H\textsubscript{2} molecule along with two charged CH\textsubscript{2} fragments.

Fig.~\ref{C4H10-snapshot} depicts the laser pulse parameters used in the Coulomb explosion simulations of 
C\textsubscript{4}H\textsubscript{10}, which included an intensity of $7 \times 10^{14}~\text{W}/\text{cm}^2$, a 
maximum electric field of $7.26~\text{V}/\text{\AA}$, a wavelength of $890~\text{nm}$, and a pulse duration of 
7.8 fs (FWHM). This intensity was determined to be optimal for achieving effective fragmentation of 
C\textsubscript{4}H\textsubscript{10}. To assess the impact of varying laser intensities on fragmentation, 
additional simulations were performed across a spectrum of intensities shown in Table~\ref{table_laser_parameters}. It was found that, similar
to the previous cases, that a particular intensity region centered
around  $7 \times 10^{14}~\text{W}/\text{cm}^2$ yielded the most
effective fragmentation.
Lower intensities resulted in insufficient dissociation, often producing larger fragments like 
C\textsubscript{2}H\textsubscript{4} without any free hydrogen dissociation, whereas higher intensities 
led to excessive dissociation, resulting in only individual atoms without any bonding. Consequently, a pulse 
intensity of $7 \times 10^{14}~\text{W}/\text{cm}^2$ was selected for the 88 simulations to ensure a diverse 
range of fragmentation pathways for C\textsubscript{4}H\textsubscript{10}.

Fig.~\ref{C4H10-histogram} displays the distribution and charges of fragments resulting from the Coulomb 
explosion of C\textsubscript{4}H\textsubscript{10}. The data indicate a diverse range of fragmentation 
products. Among these, CH\textsubscript{2} is the most frequently observed hydrocarbon, appearing in 
6.8\% of the simulations with an average charge of $0.9^{+}$. Hydrogen fragments are the most prevalent 
overall, exhibiting an average charge of $0.6^{+}$ as they approach the CAP. Other 
fragments such as CH, CH\textsubscript{2}, and C\textsubscript{2}H\textsubscript{2} are also present but occur less frequently. 
the laser pulse intensity was sufficient to remove at least three hydrogen atoms from the molecule in each simulation, as no fragments larger than C\textsubscript{3}H\textsubscript{6} were detected.  
On average, the laser ejected 5.54 electrons per simulation. These results are crucial for optimizing pulse 
parameters and predicting the fragmentation pathways of C\textsubscript{4}H\textsubscript{10} in Coulomb explosion studies.

\section{Summary}

Fragmentation in the Coulomb explosion of hydrocarbon molecules was
investigated using time dependent density functional theory simulations. This approach enables the visualization of the underlying mechanisms 
of fragment formation, the prediction of product formation pathways, and the identification of ideal 
intensities for diverse fragmentation. We have demonstrated that at specific regions of laser intensity, hydrocarbon molecules dissociate into charged fragments 
consisting of \( n \) carbon and \( m \) hydrogen atoms, denoted as C\(_n\)H\(_m\). Higher laser intensities result in the complete dissociation of the molecule, 
while lower intensities typically lead to the stripping of protons or cause only ionization (refer to Table~\ref{table_laser_parameters}). 
The C\(_n\)H\(_m\) fragments can be experimentally detected \cite{submitted}.

The optimal pulse intensities that yielded the most favorable fragmentation distributions for CH\textsubscript{4}, C\textsubscript{2}H\textsubscript{6}, C\textsubscript{3}H\textsubscript{8}, and C\textsubscript{4}H\textsubscript{10} were determined to be \(1.1 \times 10^{15}~\text{W}/\text{cm}^2\), \(9 \times 10^{14}~\text{W}/\text{cm}^2\), \(7 \times 10^{14}~\text{W}/\text{cm}^2\), and \(7 \times 10^{14}~\text{W}/\text{cm}^2\), respectively (as demonstrated in Table~\ref{table_laser_parameters}). All other laser parameters, including wavelength, duration, and center frequency, were held constant for each molecule across their respective simulations. The electric field strength—and consequently, the intensity—was the only parameter adjusted for each molecule. 

Notably, ionization of the molecule in each simulation occurred only when the laser electric field exceeded a specific range corresponding to each molecule. Fig.~\ref{CH4-snapshot}, Fig.~\ref{C2H6-snapshot}, Fig.~\ref{C3H8-snapshot}, and Fig.~\ref{C4H10-snapshot} illustrate the electric field of the laser and the number of electrons in the molecule over time, demonstrating that ionization occurs approximately 5 fs before and after the peak electric field.

The initialization of velocities based on the Boltzmann distribution at 300 K facilitated the distribution of fragment products for each molecule using the same laser pulse across all simulations. The statistical distributions of fragment production frequencies provide crucial insights into predicting the probabilities of various fragment formation pathways for each tested molecule. Furthermore, the average charge of the fragments indicates the likelihood of different charge states for a molecule. Fig.~\ref{C4H10-histogram}, for example, depicts an average charge of \(0.6^{+}\) for the produced CH species, which suggests that any CH fragment resulting from the Coulomb explosion has a 60\% probability of being positively charged and a 40\% probability of being neutral. 

Moreover, a comparison of the histograms for all molecules (Fig.~\ref{CH4-histogram}, Fig.~\ref{C2H6-histogram}, Fig.~\ref{C3H8-histogram}, Fig.~\ref{C4H10-histogram}) provides valuable insights. For instance, the average charge of the expelled free hydrogen is consistently \(0.6^{+}\) across all precursor molecules. Notable similarities are observed in the characteristics of the CH fragments as well. The two smaller alkanes (CH\textsubscript{4} and C\textsubscript{2}H\textsubscript{6}) produced CH fragments with an average charge of \(0.8^{+}\), while the two larger alkanes (C\textsubscript{3}H\textsubscript{8} and C\textsubscript{4}H\textsubscript{10}) yielded CH fragments with an average charge of \(0.6^{+}\).

The snapshot diagrams from specific simulations illustrate the formation of fragments and bonds following 
the rapid ionization induced by a strong laser field. These images provide crucial insights into the 
discharge of electron density from the molecule
and the repulsion between the resulting charged fragments. Additionally, the snapshots enable the study of 
fragment dynamics after the laser-driven ionization, including structural rearrangements 
(Fig.~\ref{CH4-snapshot}), rotations (Fig.~\ref{C2H6-snapshot}), and the transfer of atoms 
between different fragments (Fig.~\ref{C3H8-snapshot}).


Future work could involve experimental validation of these findings to refine theoretical models, particularly by exploring different 
laser parameters and molecular environments. Such experimental efforts will help bridge the gap between theory 
and practice, providing a more comprehensive understanding of fragmentation dynamics of the Coulomb explosion.

\begin{acknowledgments} 
This work has been supported by the National Science Foundation (NSF) under Grant No. 2217759.

This work used ACES at TAMU through allocation PHYS240167 from the Advanced Cyberinfrastructure Coordination Ecosystem: Services \& Support (ACCESS) program, which is supported by National Science 
Foundation grants \#2138259, \#2138286, \#2138307, \#2137603, and \#2138296~\cite{aces}.

\end{acknowledgments}

%

\end{document}